\documentclass[a4paper,11pt]{article}
\pdfoutput=1 
\usepackage{jheppub} 
\usepackage{amsmath}
\usepackage[T1]{fontenc} 
\def\bal#1\eal{\begin{align}#1\end{align}}
\def\alp[#1]{\begin{align}#1\end{align}}

\def\secnum[#1]{\texorpdfstring{$#1$}{TEXT}}

\def\secnuml#1\secnumr{\texorpdfstring{$#1$}{TEXT}}
\newcommand\beq{\begin{equation}}
\newcommand\eeq{\end{equation}}
\newcommand\bali{\begin{aligned}}
\newcommand\eali{\end{aligned}}
\def\eqa{\begin{eqnarray}}
\def\eqae{\end{eqnarray}}
\def\eq{\begin{equation}}
\def\eqe{\end{equation}}
\def\be{\begin{equation}}
\def\ee{\end{equation}}
\def\bea{\begin{eqnarray}}
\def\eea{\end{eqnarray}}
\def\ba{\begin{array}}
\def\ea{\end{array}}
\def\bd{\begin{displaymath}}
\def\ed{\end{displaymath}}

\def\>{\rangle}
\def\<{\langle}

\title{End of the World Brane meets $T\bar{T}$}
\author[a]{Feiyu Deng,}
\author[a]{Zhi Wang,}
\author[a,b]{Yang Zhou}

\affiliation[a]{Department of Physics and Center for Field Theory and Particle Physics, Fudan University, Shanghai 200433, China}

\affiliation[b]{Peng Huanwu Center for Fundamental Theory, Hefei, Anhui 230026, China}

\emailAdd{fydeng20@fudan.edu.cn}
\emailAdd{zhiwang20@fudan.edu.cn}
\emailAdd{yang\_zhou@fudan.edu.cn}

\begin{document}
\abstract{End of the world branes in AdS have been recently used to study problems deeply connected to quantum gravity, such as black hole evaporation and holographic cosmology. With non-critical tension and Neumann boundary condition, the end of the world brane often represents part of the degrees of freedom in AdS gravity and geometrically it is only part of the entire boundary. On the other hand, holographic $T\bar{T}$ deformation can also give a boundary as a cutoff surface for AdS gravity. In this paper we consider AdS gravity with both the end of the world boundary and the cutoff boundary. Using partial reduction we obtain a brane world gravity glued to a $T\bar{T}$ deformed bath. We compute both entanglement entropy and Page curve, and find agreement between the holographic results and island formula results.}



\maketitle
\flushbottom
\section{Introduction}
Solving quantum gravity is one of the most challenging problems in modern physics. In particular, it is necessary for understanding the microscopic description of black holes as well as the origin of our universe. AdS/CFT correspondence~\cite{Maldacena:1997re,Gubser:1998bc,Witten:1998qj} is a promising framework to understand quantum gravity in asymptotically AdS spacetime, which is however different from the universe we observe. It is therefore interesting to ask how to study black hole evaporation as well as cosmology evolution in the framework of AdS/CFT. In recent attempts, the end of the world (EOW) brane~\cite{Takayanagi:2011zk,VanRaamsdonk:2020ydg,Karch:2020iit} in AdS plays an important role~\cite{Almheiri:2019hni,Almheiri:2019psy,Penington:2019kki,Rozali:2019day,Sully:2020pza,Chen:2020uac,Geng:2020fxl,Geng:2021iyq,Deng:2020ent,Chu:2021gdb,Li:2021dmf,Shao:2022gpg,Deng:2022yll,Basu:2022reu,Lu:2022cgq,Verheijden:2021yrb,KumarBasak:2021rrx,Suzuki:2022xwv,Izumi:2022opi,Chang:2023gkt,Miao:2023unv,Cooper:2018cmb,Antonini:2019qkt,Chen:2020tes,Geng:2021wcq,Wang:2021xih,Miyaji:2021lcq,Antonini:2022ptt}.

With non-critical tension (in the sense of Karch-Randall~\cite{Karch:2000ct}) and Neumann boundary condition, the EOW brane often represents some part of the degrees of freedom in AdS gravity and geometrically it is only part of the entire boundary. On the other hand, holographic $T\bar T$ deformation has been intensively studied in recent years~\cite{McGough:2016lol,Giveon:2017nie,Donnelly:2018bef,Chen:2018eqk,Gorbenko:2018oov,Hartman:2018tkw,Gross:2019ach,Lewkowycz:2019xse,Jiang:2019epa,Chen:2019mis,Guica:2019nzm,Li:2020pwa,Allameh:2021moy,Araujo-Regado:2022gvw,He:2022bbb,Basu:2023bov,He:2023xnb,Apolo:2023aho,Apolo:2023vnm,Tian:2023fgf}. It can also provide a boundary as cutoff surface for AdS gravity. The boundary condition now is Dirichlet. Due to different boundary conditions, EOW brane and cutoff surface often represent quite different physics. Following the original Randall-Sundrum~\cite{Randall:1999ee}, we know that EOW brane can support a brane world gravity, while for a boundary with Dirichlet condition we should employ Maldacena duality (AdS/CFT) from our experience. The latter was recently extended to a finite cutoff surface where $T\bar T$/cut off surface correspondence has been proposed~\cite{McGough:2016lol,Hartman:2018tkw,Gross:2019ach}. In this paper we consider AdS gravity with both the end of the world boundary and the cutoff boundary. Using partial reduction~\cite{Deng:2020ent} we obtain a brane world gravity glued to a $T\bar T$ deformed bath.

There are several motivations for this construction: First, the recent holographic model of black hole evaporation based on the EOW brane always has a conformal bath. And one may wonder what if the bath becomes non-conformal. Second, our construction clarifies the difference between ``brane world holography" and AdS/CFT since two boundaries with different boundary conditions appear simultaneously. Third, the construction provides a strong support to partial reduction~\cite{Deng:2020ent}, which is a Karch-Randall reduction for only part of the AdS region between finite tension brane and zero tension brane.

This paper is organized as follows. In section~\ref{secII}, we start by constructing $T \bar T$ deformed CFT in half flat space and AdS$_2$. In both cases, the bulk dual is found to be the region between the cutoff surface and the tensionless brane. In section~\ref{secIII}, we construct the entire bulk which is bounded by the cutoff surface and the EOW brane with finite tension. Then we employ partial reduction for the bulk. We calculate the fine-grained entropy by island formula and find the agreement with defect extremal surface formula. In section~\ref{secIV}, we generalize the discussion to an evaporating black hole model. We compute the Page curve and again find the agreement between island formula and defect extremal surface formula. We also discuss the influence of $T\bar T$ deformation on Page curve. We conclude and discuss future questions in section~\ref{secV}.

\section{Holographic $T \bar{T}$ with boundary}\label{secII}
In this section, we construct the holographic dual  of $T\bar T$ deformed CFT in two types of geometry with a boundary: one is half flat space and the other is AdS$_2$. For convenience, we will call them Type I and Type II respectively. This construction will be useful later in the partial reduction of the entire bulk bounded by EOW brane and the cutoff surface. We start by introducing the holographic cutoff picture for $T\bar T$ deformation. 
\subsection{Holographic $T\bar T$}
Let us start by reviewing the cutoff picture for the holographic dual of $T\bar T$ in flat space~\cite{McGough:2016lol,Lewkowycz:2019xse}.~\footnote{The cutoff picture can be used to describe $T\bar T$ when the bulk is pure gravity~\cite{McGough:2016lol}. We will restrict ourselves to such cases.}
$T\bar{T}$ deformation of a CFT$_2$ in flat space with metric $ds^2=-dt^2+dx^2$ is defined by~\cite{Smirnov:2016lqw,Cavaglia:2016oda}
\be\label{TTbar}
\frac{dS(\lambda)}{d\lambda}=-2\pi\int d^2x\; T\bar{T}\ ,
\ee
where 
\begin{equation}\label{tto}
    T\bar{T}\equiv \frac{1}{8}\left(T^{ab}T_{ab}-(T^a_a)^2\right)
\end{equation}
with $T_{ab}=\frac{2}{\sqrt{-h}}\frac{\delta S(\lambda)}{\delta h^{ab}}$, is $T\bar{T}$ operator of the theory with Lorentzian action $S(\lambda)$. $T\bar{T}$ operator in flat space has been shown to satisfy factorization formula~\cite{Zamolodchikov:2004ce}
\begin{equation}\label{fac}
    \langle T\bar T\rangle= \frac{1}{8}\left(\langle T^{ab}\rangle\langle T_{ab}\rangle-\langle T^{a}_{a} \rangle^2\right)\ .
\end{equation} 
In order to have a bulk description with a finite cutoff, we take deformation parameter $\lambda \geq 0$.~\footnote{The bulk dual for $\lambda\leq 0$ is proposed in~\cite{Apolo:2023vnm} by a gluing procedure.} Deformed theory forms a trajectory in the $2d$ theory space. The initial condition of this trajectory is given by the seed CFT$_2$, i.e. $S(0)=S_{\mathrm{CFT}_2}$. Since $T\bar{T}$ deformation of CFT in flat space has a single mass scale $\mu=\frac{1}{\sqrt{\lambda}}$, varying the action $S(\lambda)$ with respect to $\mu$ we have $\mu\frac{dS}{d\mu}=-2\lambda\frac{dS}{d\lambda}=-\int d^2xT^a_a$. Therefore Eq.~(\ref{TTbar}) is equivalent to the trace flow equation \cite{McGough:2016lol,Kraus:2018xrn}
\be\label{tfe1}
T^a_a=-4\pi \lambda T\bar{T}\ .
\ee
Now let us consider the bulk gravity action
\begin{equation}\label{action}
I=\frac{1}{16 \pi G_{ N}} \int_{M} d^3x\sqrt{-g}\left(R+\frac{2}{l^2}\right)+\frac{1}{8 \pi G_{ N}} \int_{N} d^2x  \sqrt{-\gamma} \left(K-\frac{1}{l}\right)\ ,
\end{equation}
where $\gamma$ is the induced metric of the boundary $N$.
The holographic dual of $T\bar{T}$ deformed CFT$_2$ 
in flat space was proposed to be the quantum gravity in AdS$_3$ with a radial cutoff in Poincare coordinate~\cite{McGough:2016lol,Lewkowycz:2019xse}
\be\label{bmet1}
ds^2=\frac{l^2}{z^2}\left( dz^2-dt^2+dx^2\right)\;\;\; ,\;\;\;
z\geq z_c\ ,
\ee
where the Dirichlet boundary condition is imposed at $z=z_c$.
The $T\bar T$ deformed theory lives on a fixed background
\begin{equation}\label{met1}
    ds^2=-dt^2+dx^2\ ,
\end{equation}
with the deformation parameter related to the radial cutoff by~\cite{Lewkowycz:2019xse}
\be\label{dic1}
c=\frac{3l}{2G_N}\ ,\;\;\;
\lambda=\frac{8G_N}{l}z_c^2=\frac{12}{c}z_c^2\ .
\ee
If we rescale the background metric to the induced metric of the cutoff surface
\be
ds^2=\frac{l^2}{z_c^2}(-dt^2+dx^2)\ ,
\ee
and simultaneously rescale $\lambda$ to $\frac{z_c^2}{l^2}\lambda$, the defining equation~(\ref{TTbar}) is still satisfied, therefore one can equivalently consider the $T\bar T$ deformed theory on the cutoff surface but with a rescaled holographic dictionary~\cite{Kraus:2018xrn}
\be\label{dica}
c=\frac{3l}{2G_N}\;\;\;,\;\;\;
\lambda=8G_Nl=\frac{12l^2}{c}\ .
\ee

Now we generalize the cutoff picture to $T\bar T$ deformed CFT in AdS$_2$. Notice that in large central charge limit, the $T\bar{T}$ deformation
\be\label{ttb2}
\frac{dS(\lambda)}{d\lambda}=-2\pi\int d^2x\sqrt{-h}\; T\bar{T}
\ee
is well-defined~\cite{Jiang:2019tcq,Brennan:2020dkw} for CFT in AdS$_2$ with Poincare metric $h_{ab}$
\begin{equation}\label{bcm2}
    ds^2=\frac{l^2}{u^2}\left(-dt^2+du^2\right)\ ,
\end{equation}
and it still satisfies the factorization formula Eq.~(\ref{fac}). In this case, the trace flow equation including curvature contribution becomes~\cite{McGough:2016lol,Shyam:2017znq}
\be\label{tfe2}
T^a_a=-\frac{c}{24\pi}\mathcal{R}[h]-4\pi \lambda T\bar{T}\ ,
\ee
where $\mathcal{R}[h]=-\frac{2}{l^2}$ is Ricci scalar of the AdS$_2$ space. 

To figure out the holographic dual, we take AdS$_2$ slicing of AdS$_3$ space
\begin{equation}
    d s^{2}
=d \rho^{2}+l^2\cosh ^{2} \frac{\rho}{l} \left(\frac{-d t^{2}+d u^{2}}{u^{2}}\right)\ .
\end{equation}
In these coordinates, AdS/CFT duality implies that there are two dual CFTs~\cite{Aharony:2010ay,Ghodsi:2022umc} on the two asymptotic boundaries $\rho\rightarrow\pm \infty$. Then at finite cutoff surfaces, two $T\bar T$ deformed CFTs with the same deformation parameter $\lambda$ are dual to the finite wedge region 
\be\label{poincareslice}
ds^2=d\rho^2+l^2\cosh^2\frac{\rho}{l}\left(\frac{-dt^2+du^2}{u^2}\right)\;\;\;,\;\;\;-\rho_c\leq\rho\leq \rho_c\ ,
\ee
where $\rho=\pm\rho_c$ are two AdS$_2$ boundaries and we impose Dirichlet boundary conditions for both.\footnote{Notice that although the wedge region looks the same as the one in wedge holography~\cite{Akal:2020wfl}, boundary conditions are different.}
To obtain the dictionary for parameters, we compare Brown-York tensor to the trace flow equation ~(\ref{tfe2}). The extrinsic curvature on the Dirichlet boundary $\rho=\rho_c$ is 
\begin{equation}
    K_{ab}=\frac{\tanh{\frac{\rho_c}{l}} }{l}\gamma_{ab}=\frac{\tanh{\frac{\rho_c}{l}} }{l}\begin{bmatrix} 

    -\frac{l^2\cosh ^{2} \frac{\rho_c}{l}}{u^2} & 0 \\

      0 & \frac{l^2\cosh ^{2} \frac{\rho_c}{l}}{u^2}

      \end{bmatrix}\ .
\end{equation}
The Brown-York tensor $T_{ab}^{\text{BY}}=\frac{1}{8 \pi G_N}\left(K_{ab}-K \gamma_{ab}+\frac{1}{l} \gamma_{ab}\right)$ is given by 
\begin{equation}\label{BY}
    T_{ab}^{\text{BY}}=-\frac{1}{8 \pi G_N}\frac{1-\tanh{\frac{\rho_c}{l}}}{l}\begin{bmatrix} 

    -\frac{l^2\cosh ^{2} \frac{\rho_c}{l}}{u^2} & 0 \\

      0 & \frac{l^2\cosh ^{2} \frac{\rho_c}{l}}{u^2}

      \end{bmatrix}\ .
\end{equation}
Given the definition of $T \bar{T}$ operator in Eq.~(\ref{tto}), the trace of Brown-York tensor satisfies
\begin{equation}\label{TF}
    (T_a^a)^{\text{BY}}=\frac{1}{8\pi G_N l}\frac{1}{\cosh^2{\frac{\rho_c}{l}}}-32\pi G_N l (T\Bar{T})^{\text{BY}}.
\end{equation}
Since the dual field theory lives on fixed background $h_{ab}=\frac{1}{\cosh^2\frac{\rho_c}{l}}\gamma_{ab}$, the Brown-York tensor can be transformed to field theory stress tensor~\cite{Hartman:2018tkw} by $T^{\text{BY}}_{ab}=\cosh^{(2-d)}\frac{\rho_c}{l} T_{ab}=T_{ab}$. Eventually the trace flow equation becomes
\begin{equation}
     (T_a^a)^{\text{BY}}=\frac{1}{\cosh^2\frac{\rho_c}{l}}T_{a}^{a}=\frac{1}{8\pi G_N l}\frac{1}{\cosh^2{\frac{\rho_c}{l}}}-\frac{32\pi G_N l}{\cosh^4{\frac{\rho_c}{l}}} T\bar{T}\ .
\end{equation}
By comparing it with CFT trace flow equation Eq.~(\ref{tfe2}), we obtain the dictionary
\begin{equation}\label{dic2}
    c=\frac{3l}{2G_N}\;\;\;,\;\;\;\lambda=\frac{8G_Nl}{\cosh^2{\frac{\rho_c}{l}}}=\frac{12l^2}{c}\frac{1}{\cosh^2\frac{\rho_c}{l}}\ .
\end{equation}
We can also rescale the background metric (\ref{bcm2}) to the induced metric of the cutoff surface and simultaneously rescale $\lambda$ to $\frac{1}{\cosh^2\frac{\rho_c}{l}}\lambda$. The rescaled holographic dictionary is the same as Eq.~(\ref{dica}).
Notice that the above discussion is on-shell, and one can also use off-shell method~\cite{McGough:2016lol,Kraus:2018xrn,Gorbenko:2018oov} to derive the same dictionary. Below in~Sec.~\ref{EE}, we compute entanglement entropy (EE) and find agreement between field theory result and holographic result, which strongly supports the duality.

\subsection{$T\bar T$ with boundary}\label{EE}

If the $T\bar T$ deformed theory in flat or AdS space preserves space-time $Z_2$ symmetry with fixed point $x=0$ or $u=0$, one can then take $Z_2$ quotient to obtain $T\bar T$ deformed theory with a boundary. In particular, in the holographic bulk, the $Z_2$ quotient will lead to a tensionless EOW brane.

\subsubsection{Type I: $T\bar T$ in half flat space}
\begin{figure}
	\centering
	\includegraphics[width=12cm,height=3.7cm]{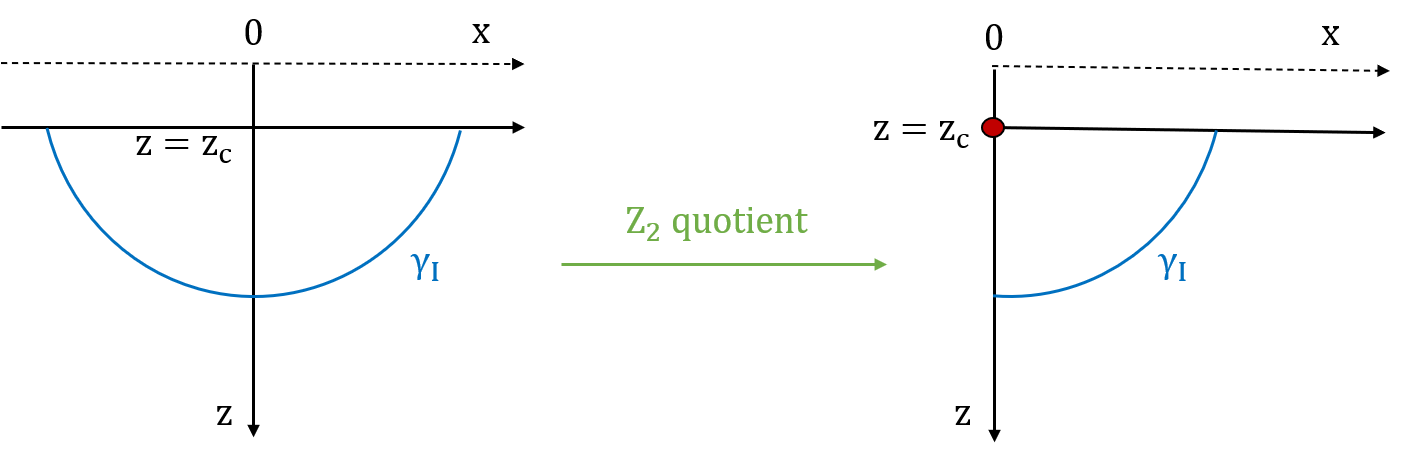}\\
	\caption{\label{Z21} Type I $T\bar{T}$ deformed theory from $Z_2$ quotient.}
\end{figure}

The holographic dual of $2d$ $T \bar{T}$ deformation in flat space is given by a radial cut off at $z=z_c$ in $3d$ AdS.
By taking $Z_2$ quotient, one can obtain $T \bar{T}$ deformed theory with a boundary along $x=0$. The dual bulk is given by
\be
ds^2=\frac{l^2}{z^2}\left(dz^2-dt^2+dx^2\right)\;\;\;,\;\;\;
z\geq z_c\;\;\;,\;\;\;x\geq 0\ .
\ee
This is equivalent to consider a tensionless EOW brane at $x=0$~\cite{Aharony:2010ay,Kastikainen:2021ybu}. The $Z_2$ quotient procedure is shown in F.G.~\ref{Z21}.
Notice that although the boundary of the deformed bath still sits at $x=0$, it moves into the bulk from $z=\epsilon$ to $z=z_c$ holographically. 

\begin{figure}
	\centering
	\includegraphics[width=10.32cm,height=4.10cm]{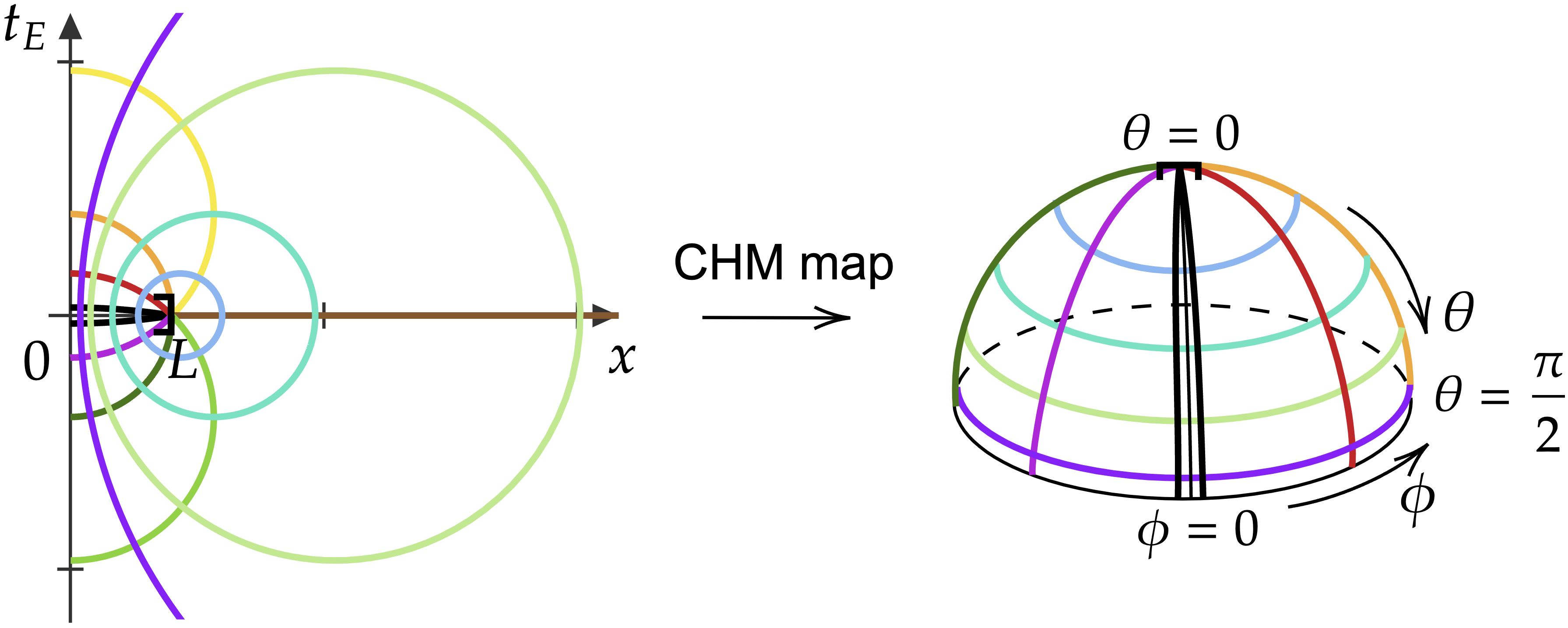}\\
	\caption{\label{CHM} Coordinate relation of CHM map from half flat space to hemisphere with radius $L$. We mark the constant $\theta$ and $\phi$ slices of the hemisphere and their related slices in half flat space by different colors. The $n$-replica space is obtained by a cyclic gluing of $n$ copies of half flat space along the cut $x\in[0,L]$. After the CHM map it corresponds to a $n$-replica hemisphere which is obtained by a cyclic gluing of $n$ copies of hemisphere along $\phi=0$.}
\end{figure}

Let us now compute the $C$-function from both sides to support the duality. By replica trick~\cite{Calabrese:2004eu}, the vacuum EE of interval $t=0,x\in[0, L]$ in half flat space $x\geq 0$ is expressed as
\begin{equation}
S(L)=\lim_{n\rightarrow1}S_n(L)=\lim_{n\rightarrow1}(1-n\partial_n)\log Z_n\ ,
\end{equation}
where $S_n$ represents $n$-Renyi entropy and $Z_n$ is the partition function of the $T\bar T$ deformed theory in the $n$-replica half flat space. The  $C$-function can be computed as $C(L)\equiv L S'(L)$ following the approach in~Ref.~\cite{Lewkowycz:2019xse}. Since the partition function $Z_n$ changes with the overall scale $L$ as $L\frac{d \log Z_n}{d L}=-\int d^2x\sqrt{g}\langle T^a_a\rangle_n$, one can express the $C$-function as
\begin{equation}
    C(L)=\lim_{n\rightarrow1}LS'_n(L)=\lim_{n\rightarrow1}\left(1-n\partial_n\right)L\frac{d \log Z_n}{d L}=-\lim_{n\rightarrow1}\left(1-n\partial_n\right)\int d^2x\sqrt{g}\langle T^a_a\rangle_n\ ,
\end{equation}
where $\langle T^a_a\rangle_n$ is the expectation value of the trace of the stress tensor in the $n$-replica half flat space. As shown in~\cite{Lewkowycz:2019xse}, the contribution of $\partial_n\langle T^a_a\rangle_n$ to $C(L)$ is localized on the endpoint of the interval
\begin{equation}
    C(L)=L S'(L)=\lim_{n\rightarrow1}(2\pi n)\int_{0}^{\rho_0\ll L}\rho d\rho \;n\partial_{n}\langle T^a_a\rangle_n\ ,
\end{equation}
where $\rho$ is the radial coordinate around the endpoint. To obtain $\partial_n\langle T^a_a\rangle_n$ near the endpoint,
we consider the CHM map~\cite{Casini:2011kv} 
\begin{equation}
t_E=\frac{L\sin\theta\sin\phi}{1+\sin\theta\cos\phi}\;\;\;,\;\;\;x=\frac{L\cos\theta}{1+\sin\theta\cos\phi}\ ,  
\end{equation}
which maps the $n$-replica half flat space to the $n$-replica hemisphere up to a conformal factor
\begin{equation}
ds^2_n=dt_E^2+dx^2=\Omega_n(\theta,\phi)^2L^2(d\theta^2+n^2\sin^2\theta d\phi^2)\ ,
\end{equation}
where $\Omega_n(\theta,\phi)=\frac{1}{1+\sin\theta\cos\phi}$. We show the map with $n=1$ in F.G.~\ref{CHM}, and for general $n$ we just map each copy of half flat space to hemisphere and glue them cyclically.
After the map, the $C$-function can be written as~\cite{Lewkowycz:2019xse}
\begin{equation}
    C(L)=L S'(L)=\lim_{n\rightarrow1}(2\pi n)L^2\int_{0}^{\theta_0\ll 1} d\theta\;\sin\theta n\partial_{n}\langle T^a_a\rangle_{{\text{HS}_n^2},\,\Tilde{\lambda }=\Omega_n^{-2}\lambda}\ ,
\end{equation}
where the expectation value of $T_a^a$ is computed from the $T\bar T$ deformed CFT in $n$-replica hemisphere $\text{HS}_n^2$ with a spacetime-dependent deformation parameter $\tilde{\lambda}$. However, since the conformal factor is 1 at the endpoint $\theta=0$, $\tilde{\lambda}$ equals to $\lambda$ at the endpoint. Thus near the endpoint 
\begin{equation}\label{HST}
    n\partial_{n}\langle T^a_a\rangle_{{\text{HS}_n^2},\,\tilde{\lambda }}=n\partial_{n}\langle T^a_a\rangle_{{\text{HS}_n^2},\,\lambda}\ .
\end{equation} 
Furthermore, the $Z_2$ quotient boundary condition of replica hemisphere relates (\ref{HST}) to the one on entire replica sphere $\text{S}_n^2$. The stress tensor for $T\bar T$ deformed CFT in $\text{S}_n^2$ has been obtained by solving the trace flow equation and conservation equation in~\cite{Donnelly:2018bef}.
Then using Eq.~(3.10) and (3.11) of~\cite{Donnelly:2018bef}, we obtain the $C$-function for $T\bar T$ deformed CFT in half flat space
\begin{equation}\label{cf1b}
    C(L)=\frac{c}{6}\frac{1}{\sqrt{1+\frac{\lambda c}{12L^2}}}\ .
\end{equation}
We can also compute the $C$-function from the bulk. As shown in F.G.\ref{Z21}, 
in the bulk EE is given by the area of RT surface from cutoff boundary to zero tension brane 
\begin{equation}\label{npe1}
    S(L)=\frac{\text{Area}(\gamma_I)}{4G_N}=\frac{c}{6}\log\left(\frac{L}{z_c}+\sqrt{1+\left(\frac{L}{z_c}\right)^2}\right)\ .
\end{equation}
By the definition $C(L)=LS'(L)$ and the dictionary Eq.~(\ref{dic1}), we can see that it gives the same $C$-function as Eq.~(\ref{cf1b}).

\subsubsection{Type II: $T\bar T$ in AdS$_2$}\label{II}

\begin{figure}
	\centering
	\includegraphics[width=15cm,height=3.5cm]{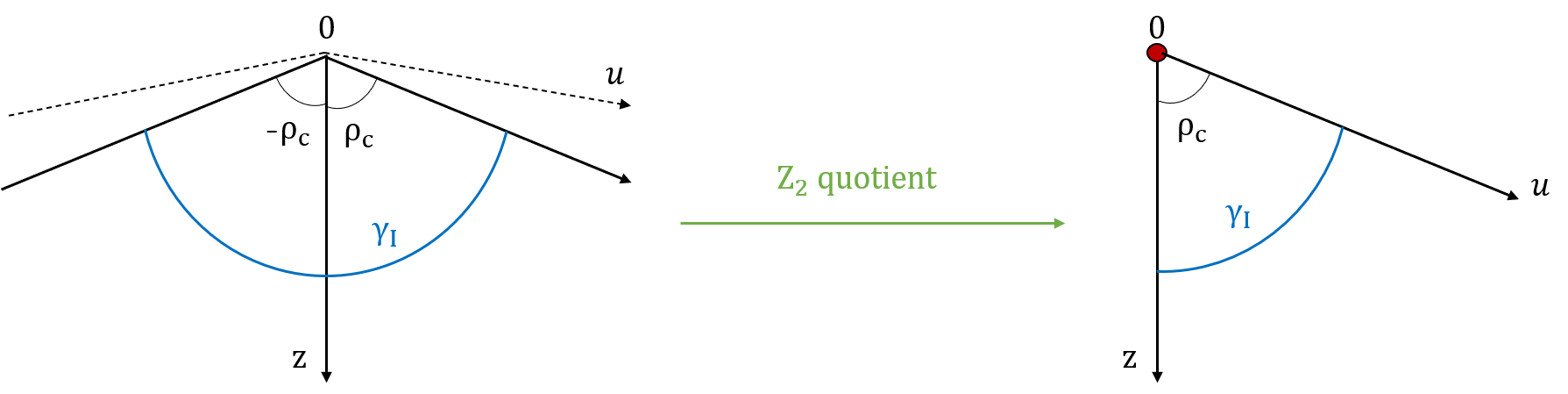}\\
	\caption{\label{Z22} Type II $T\bar{T}$ deformed CFT bath with zero boundary entropy from $Z_2$ quotient.}
\end{figure}

 In our proposed duality between $T\bar T$ deformed CFT in AdS$_2$ and the corresponding wedge, two $T\bar T$ deformed CFT have the same deformation parameter. The field theory background is $Z_2$ invariant along $u=0$ and the bulk dual is $Z_2$ invariant along $\rho=0$. After $Z_2$ quotient, we obtain $T\bar T$ deformed CFT in a single AdS$_2$ background with an asymptotic boundary at $u=0$. The bulk dual after $Z_2$ quotient is given by
\be\label{IIdual}
ds^2=d\rho^2+l^2\cosh^2\frac{\rho}{l}\left(\frac{-dt^2+du^2}{u^2}\right)\;\;\;,\;\;\;0\leq\rho\leq \rho_c\ .
\ee
Again the bulk $Z_2$ fixed point corresponds to a tensionless EOW brane~\cite{Aharony:2010ay,Andrade:2011nh}.
The $Z_2$ quotient procedure to obtain Type II $T\bar T$ deformed CFT is shown in F.G.\ref{Z22}.

\begin{figure}
	\centering
	\includegraphics[width=11.68cm,height=4.17cm]{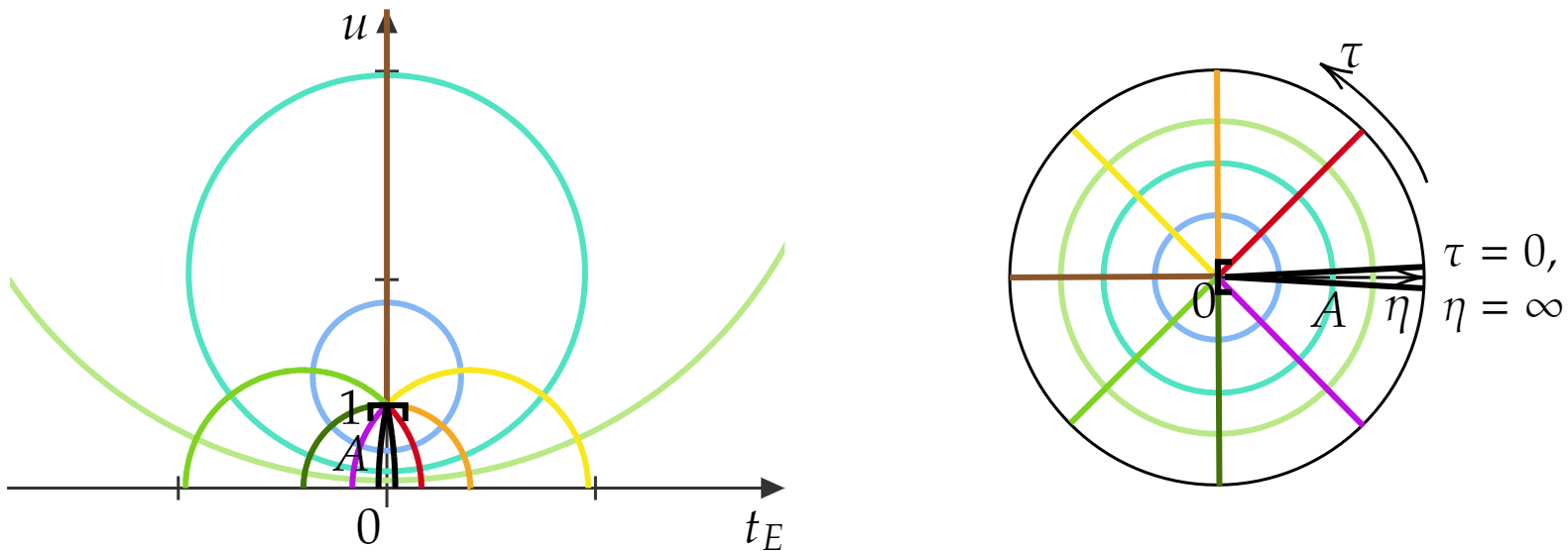}\\
	\caption{\label{PHH} Coordinate transformation between Euclidean Poincare AdS$_2$ coordinate $(t_E,u)$ in left figure and Hartle-Hawking coordinate $(\eta,\tau)$ in right figure. For interval $A$, the $n$-replica space is obtained by cyclic gluing of $n$ copies of $\mathbb{H}^2$ along the cut $\tau=0$.}
\end{figure}

Now we calculate EE of a spatial interval on both sides to support the duality. For Type II case, we consider EE of an interval $u\in [0,u_0]$ at $t=0$ in Poincare vacuum. As discussed in~\cite{Spradlin:1999bn}, AdS$_2$ Poincare vacuum is equivalent to the global vacuum and the Hartle-Hawking vacuum. By analytic continuation to Euclidean signature, these vacuum states are defined by 
Euclidean path integral on the hyperbolic space $\mathbb{H}^2$. The metric of $\mathbb{H}^2$ used to define these vacuum states is
\begin{equation}
\begin{split}
ds_{\mathbb{H}^2}^2
        &=l^2\frac{dt_E^2+du^2}{u^2}\\
        &=l^2 \frac{4dwd\bar{w}}{(1-w\bar{w})^2}\\
        &=l^2\left(d\eta^2+\sinh^2\eta d\tau^2\right)\ ,\\
\end{split}
\end{equation}
where $t_E=-i t$ is Euclidean time and we perform coordinate transformation to complex Poincare disk $(w,\bar w)$ by $w=\frac{1-u-it_E}{1+u+it_E}$. Hartle-Hawking coordinate $(\eta,\tau)$ is related to Poincare coordinate by 
\begin{equation}
    t_E=\frac{\sinh\eta\sin\tau}{\cosh\eta+\cos\tau\sinh\eta}\;\;\;,\;\;\;
    u=\frac{1}{\cosh\eta+\cos\tau\sinh\eta}\ .
\end{equation}
We show this coordinate transformation in F.G.~\ref{PHH}.
In complex Poincare disk coordinates, the interval becomes $w\in[\frac{1-u_0}{1+u_0},1]$. Below we consider a special interval $A$ with $w\in[0,1]$, which 
 corresponds to $u_0\in[0,1]$ or $\eta_0\in[0,\infty)$, as shown in F.G.~\ref{PHH}. For this special interval, we can easily obtain the $n$-replica manifold in Hartle-Hawking coordinate with the metric
\begin{equation}\label{rep}
ds^2_{\mathbb{H}^2_n}=l^2\left(d\eta^2+n^2\sinh^2\eta d\tau^2\right)
\end{equation}
and we follow the procedure in Ref.~\cite{Donnelly:2018bef} to compute EE in the $T\bar T$ deformed vacuum.

Consider $T\bar{T}$ deformed CFT in hyperbolic space
\begin{equation}
ds^2_{\mathbb{H}^2}=l^2\left(d\eta^2+\sinh^2\eta d\tau^2\right)\ ,
\end{equation}
its partition function $\log Z$ varies with respect to AdS radius $l$ as
\begin{equation}\label{pz}
\begin{split}
        \frac{d}{d l}\log Z&=2l\frac{\partial \log Z}{\partial h_{\eta\eta}}+2l\sinh^2\eta\frac{\partial \log Z}{\partial h_{\tau\tau}}\\
        &=-\int d^2x \sqrt{h}\left(l\langle T^{\eta\eta}\rangle+l\sinh^2\eta \langle T^{\tau\tau}\rangle \right)\\
        &=-\frac{1}{l}\int d^2x \sqrt{h}\langle T^a_a\rangle\ ,
\end{split}
\end{equation}
where we use the definition $\int d^2x\sqrt{h} \langle T^{ab}\rangle=-2\frac{\partial \log Z}{\partial h_{ab}}$. Since $\mathbb{H}^2$ is maximally symmetric space, the one-point function of stress tensor is proportional to metric $\langle T_{ab}\rangle=\alpha h_{ab}$. Substituting into trace flow equation Eq.~(\ref{tfe2}), we obtain
\begin{equation}\label{st}
    \langle T_{ab}\rangle=\frac{1}{\pi\lambda}\left(1-\sqrt{1-\frac{c\lambda}{12 l^2}}\right)h_{ab}\;\;\;,\;\;\;\langle T_{a}^{a}\rangle=\frac{2}{\pi\lambda}\left(1-\sqrt{1-\frac{c\lambda}{12 l^2}}\right)\ .
\end{equation}
Notice that when the deformation parameter $\lambda>\lambda_c\equiv\frac{12 l^2}{c}$, the spectrum will include imaginary energy. By using the dictionary Eq.~(\ref{dic2}) we find that the critical value $\lambda_c$ corresponds to the cutoff surface $\rho_c=0$, which means that all the spectrum in our consideration is real. Using expression of $\langle T_{a}^{a}\rangle$, Eq.~(\ref{pz}) becomes
\begin{equation}\label{dllogz}
    \frac{d}{d l}\log Z=\frac{4}{\lambda}\left(l-\sqrt{l^2-\frac{c \lambda}{12}}\right)\ ,
\end{equation}
where we have used the renormalized volume of unit hyperbolic space $V_{\mathbb{H}_2}=-2\pi$. We can integrate this differential equation with respect to $l$ and obtain
\begin{equation}\label{pt2}
\log Z=\frac{c}{6} \log \left[\frac{l}{a}\left(1+\sqrt{1-\frac{c \lambda}{12 l^2}}\right)\right]-\frac{2 l^2}{\lambda} \sqrt{1-\frac{c \lambda}{12 l^2}}+\frac{2 l^2}{\lambda}\ ,
\end{equation}
where $a$ is an arbitrary integration constant with length dimension.
The physical meaning of $a$ is the finite cutoff of the deformed theory which depends on the deformation parameter~\cite{Apolo:2023vnm}. To determinate $a(\lambda)$, we substitute Eq.~(\ref{pt2}) into another differential equation~\footnote{This is equivalent to the classical definition (\ref{ttb2}) of $T\bar T$ deformation, up to a Weyl anomaly term.} 
\begin{equation}\label{df}
\mu\partial \mu \log Z=-2\lambda\partial_{\lambda} \log Z=-\int d^2x \sqrt{h}\langle T^a_a\rangle=-\int d^2x\sqrt{h}\left(\frac{c}{24\pi}\mathcal{R}[h]-4\pi\lambda \langle T\bar T\rangle\right)\ ,
\end{equation}
which is the equation of $\log Z$ varying with respect to the deformation scale $\mu=\frac{1}{\sqrt{\lambda}}$. We take the boundary condition at the critical point to be~\footnote{Below we can find that this boundary condition is equivalent to EE $S(A)=0$ when $\rho_c=0$.} 
\begin{equation}
   \log Z|_{\lambda=\lambda_c}=-I_E|_{\rho_c=0}=-\frac{1}{8\pi G_N}\int_{\rho=0}d^2x\sqrt{\gamma}\frac{1}{l}\\=\frac{l}{4 G_N}=\frac{c}{6}\ , 
\end{equation}
which gives $a=\sqrt{\frac{c\lambda}{12}}$. As a consistent check, we evaluate the bulk Euclidean on-shell action~(\ref{action}) of the dual geometry~(\ref{IIdual}) at general $\rho_c$
\begin{equation}\label{acb}
\begin{split}
        I_E&=-\frac{1}{16\pi G_N}(-2\pi) \int_{0}^{\rho_c}d\rho \;l^2\cosh^2\frac{\rho}{l}(-\frac{4}{l^2})-\frac{1}{8\pi G_N}(-2\pi)l^2\cosh^2\frac{\rho_c}{l}(\frac{2}{l}\tanh\frac{\rho_c}{l}-\frac{1}{l})\\
        &=-\frac{l}{8G_N}\left(1+e^{-\frac{2\rho_c}{l}}+\frac{2\rho_c}{l}\right)\ ,
\end{split}
\end{equation}
we find that it exactly equals to $-\log Z$ in Eq.~(\ref{pt2}) with $a=\sqrt{\frac{c\lambda}{12}}$ and (\ref{dic2}).~\footnote{We can check that the term proportional to $\frac{2\rho_c}{l}$ in Eq.~(\ref{acb}) is the $\log$ divergence term in Fefferman-Graham coordinate, and it is just the first term in Eq.~(\ref{pt2}).}

Now we use the partition function that we have obtained to compute EE. The first derivative with respect to $n$ of $\log Z_n$ at $n=1$ is
\begin{equation}
\begin{split}
    \frac{d \log Z_n}{dn}\bigg|_{n=1}
    &=-\int d^2x\sqrt{h}\;2L^2\sinh^2\eta \frac{\partial \log Z_n}{\partial h_{\tau\tau}}\bigg|_{n=1}\\
    &=-\int d^2x\sqrt{h}\;\langle T^{\tau}_{\tau}\rangle\\
    &=-\frac{1}{2}\int d^2x\sqrt{h}\;\langle T^a_a\rangle\ .
\end{split}
\end{equation}
From this, we obtain the EE
\begin{equation}
\begin{split}
    S(A)&=\left(1-n\frac{d}{dn}\right)\log Z_n\bigg|_{n=1}\\
    &=\left(1-\frac{l}{2}\frac{d}{dl}\right)\log Z\\
    &=\frac{c}{6}\log\frac{l\left(1+\sqrt{1-\frac{c\lambda}{12 l^2}}\right)}{\sqrt{\frac{c\lambda}{12}}}\\
    &=\frac{c}{6}\frac{\rho_c}{l}\ ,
\end{split}
\end{equation}
where we have used dictionary Eq.~(\ref{dic2}).
The $C$-function is calculated as
\begin{equation}
    C=l\frac{dS}{dl}=\frac{l}{2}\frac{d\log Z}{dl}-\frac{l^2}{2}\frac{d^2\log Z}{dl^2}=\frac{c}{6}\frac{1}{\sqrt{1-\frac{c\lambda}{12 l^2}}}
    =\frac{c}{6}\coth\frac{\rho_c}{l}\ .
\end{equation}
We find that $C$-function is a monotonic decreasing function which becomes $\log$ coefficients $\frac{c}{6}$ of BCFT EE when $\rho_c\rightarrow\infty$ and diverges when $\rho_c\rightarrow0$.

We can also calculate EE by RT formula. As shown in F.G.\ref{Z22}, the geodesic length from $\rho=\rho_c$ to tensionless brane at $\rho=0$ is a constant $\gamma_{I}=\rho_c$, so the holographic EE of interval $A$ is given by
\begin{equation}\label{npe2}
    S(A)=\frac{\text{Area}(\gamma_I)}{4G_N}=\frac{\rho_c}{4G_N}=\frac{c}{6}\frac{\rho_c}{l}\ .
\end{equation}
We see that this holographic EE exactly matches field theory result at non-perturbative level.

\section{EOW brane meets $T \bar{T}$}\label{secIII}

In this section, we construct the AdS bulk with both EOW brane and finite cutoff Dirichlet boundary. Using partial reduction, we obtain the brane world gravity coupled with $T\bar T$ deformed bath. To maintain a transparent boundary condition we also consider a $T\bar T$ deformed defect theory on the brane. We compare fine-grained entropy from defect extremal surface (DES) and from island formula, and find agreement.

\subsection{Bulk construction}\label{I}

Now we construct the bulk solution with both EOW brane and holographic $T\bar T$ cutoff boundary. The brane and the boundary intersect with each other. The shape and position of EOW brane can be solved from equation of motion and for simplicity we take the brane tension to be a constant $T$.

\begin{figure}
	\centering
	\includegraphics[width=14cm,height=4.5cm]{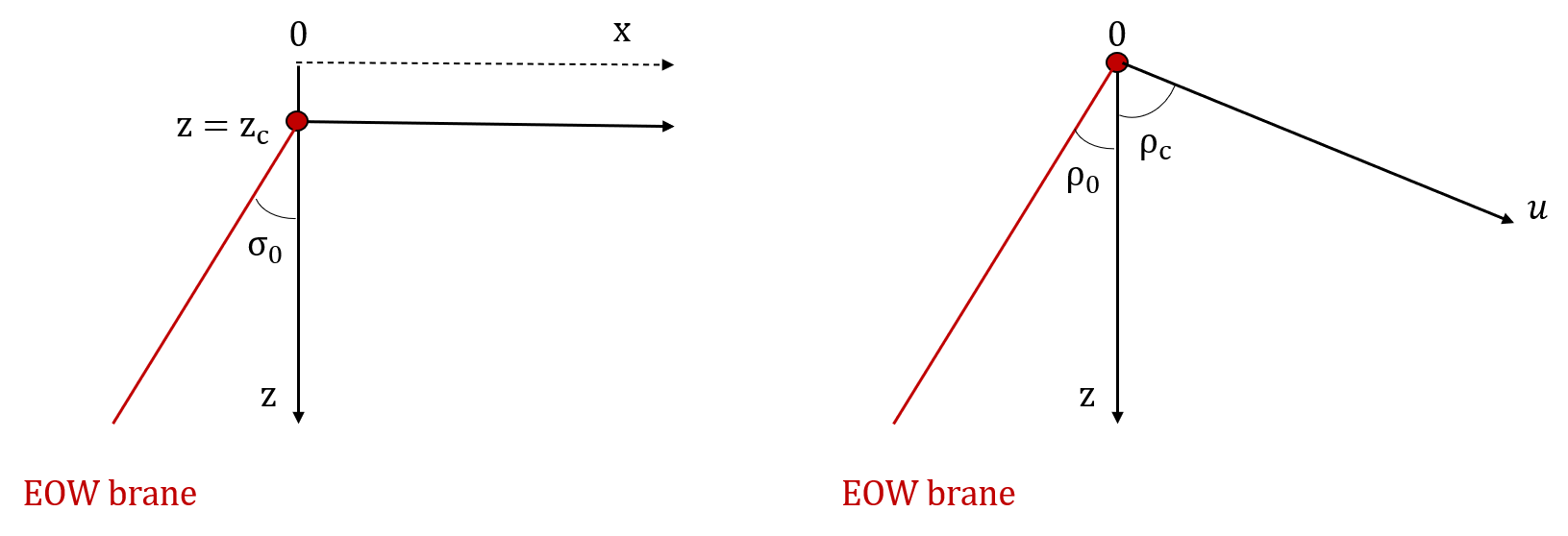}\\
	\caption{Bulk construction for Type I deformed bath (left) and for Type II (right). }\label{n1}
\end{figure}

We first construct the bulk for Type I case. Notice that in the bulk, the boundary of $Z_2$ quotient $T\bar T$ CFT moves from $z=\epsilon$ to $z=z_c$, therefore the EOW brane should starts from $z=z_c$. Let us consider bulk coordinate transformations
\begin{equation}
    \begin{split}
        z&=z_c+\frac{y}{\cosh{\frac{\sigma}{l}}}\ ,\\
        x&=-y\tanh{\frac{\sigma}{l}}\ ,
    \end{split}
\end{equation}
then the bulk metric becomes
\begin{equation}
    \begin{split}
        ds^2&=\frac{l^2}{z^2}(-dt^2+dx^2+dz^2)\\
        &=\frac{l^2}{(z_c+\frac{y}{\cosh{\frac{\sigma}{l}}})^2}(-dt^2+dy^2+\frac{y^2}{l^2\cosh^2{\frac{\sigma}{l}}}d\sigma^2)\ .
    \end{split}
\end{equation}
The normal vector to $\sigma=\sigma_0$ slice is
\begin{equation}
    n_a=\frac{y}{\cosh{\frac{\sigma_0}{l}}(z_c+\frac{y}{\cosh{\frac{\sigma_0}{l}}})}(0,0,1)\ ,
\end{equation}
thus the extrinsic curvature $K_{ab}=\nabla_a n_b$ is given by
\begin{equation}
    K_{ab}=\frac{\tanh{\frac{\sigma_0}{l}}}{l}\begin{bmatrix} 
    -\frac{l^2}{(z_c+\frac{y}{\cosh{\frac{\sigma}{l}}})^2} & 0 \\

      0 & \frac{l^2}{(z_c+\frac{y}{\cosh{\frac{\sigma}{l}}})^2}
      \end{bmatrix}\ .
\end{equation}
Plugging into Neumann boundary condition on EOW brane
\begin{equation}\label{NBC}
    K_{ab}-h_{ab}(K-T)=0\ ,
\end{equation}
we find that $\sigma=\sigma_0$ is a solution with the tension $T=\frac{\tanh{\frac{\sigma_0}{l}}}{l}$. Bulk construction for Type I is shown in F.G.\ref{n1}.

Now let us consider bulk construction in Type II case. Notice that from bulk point of view the boundary position of cutoff surface remains the same under deformation. Therefore the shape and position of EOW brane stays the same as if there is no $T\bar{T}$ deformation. They are determined by solving Neumann boundary condition (\ref{NBC}).
The result is that EOW brane lies along $\rho=\rho_0$ slice and the relation between $\rho_0$ and tension $T$ is $T=\frac{\tanh{\frac{\rho_0}{l}}}{l}$.
See F.G.\ref{n1} for an illustration.

After solving EOW brane configuration, now we would like to include the defect theory on the brane. Since we want to impose transparent boundary condition for the $T\bar T$ deformed bath, we include the same field theory on the EOW brane which has the same deformation parameter as the $T\bar T$ deformed CFT on the cutoff boundary. 
The Neumman boundary condition with defect matter is
\begin{equation}
K_{a b}-\gamma_{a b}(K-T)=8 \pi G_N   \langle T_{a b}\rangle\ .
\end{equation}
Taking trace we have
\begin{equation}\label{nec1}
    2T-K=8 \pi G_N  \langle T^{a}_{a }\rangle\ ,
\end{equation}
where $T_{ab}$ is the stress tensor of the defect matter.
We assume the new EOW brane is still along a constant $\omega_1$ slice , where $\omega_1=\sigma_0$ for Type I and $\omega_1=\rho_0$ for Type II. Then the induced metric is an AdS$_2$ with radius $l\cosh\frac{\omega_1}{l}$.
Notice that the stress tensor for $T\bar T$ deformed CFT in AdS$_2$ is given by Eq.~(\ref{st}). Then the Neumman boundary condition becomes
\begin{equation}\label{re}
    2T-\frac{2}{l}\tanh{\frac{\omega_1}{l}}=8\pi G_N\frac{2}{\pi \lambda}\left(1-\sqrt{1-\frac{c\lambda}{12l^2\cosh^2(\frac{\omega_1}{l})}}\right)=\frac{2}{l}(1-\tanh\frac{\omega_1}{l})\ ,
\end{equation}
where in the second equality we use the dictionary (\ref{dica}). Then the Neumman boundary condition fixes the classical tension $T=\frac{1}{l}$ and $\omega_1$ can be arbitrary positive constant. This consequence is due to the holographic nature of $T\bar T$ CFT on the brane. If it is non-holographic, then we can not use dictionary (\ref{dica}) in Eq. (\ref{re}) and the position of brane $\omega_1$ will generically depend on tension $T$ and deformation parameter $\lambda$.

\subsection{Partial reduction}
After the bulk is constructed, we now discuss how to obtain the effective description where Island formula~\cite{Almheiri:2019hni} is applicable. This can be done by employing the so-called partial reduction~\cite{Deng:2020ent}, i.e. doing dimensional reduction for the wedge between EOW brane and zero tension brane, while using the duality we have developed in section~\ref{secII} for the rest of the bulk. 
We dualize the rest of the bulk to $T\bar T$ deformed theory on the cutoff surface, which means we take the dictionary~(\ref{dica}). Eventually we obtain an effective theory including a gravity theory coupled with $T\bar T$ deformed CFT matter on EOW brane, and also a $T\bar T$ deformed CFT bath glued to EOW brane. In this effective theory, one can use island formula to compute fined-grained entropy for an interval in the bath. 

\begin{figure}
	\centering
	\includegraphics[width=14cm,height=4.2cm]{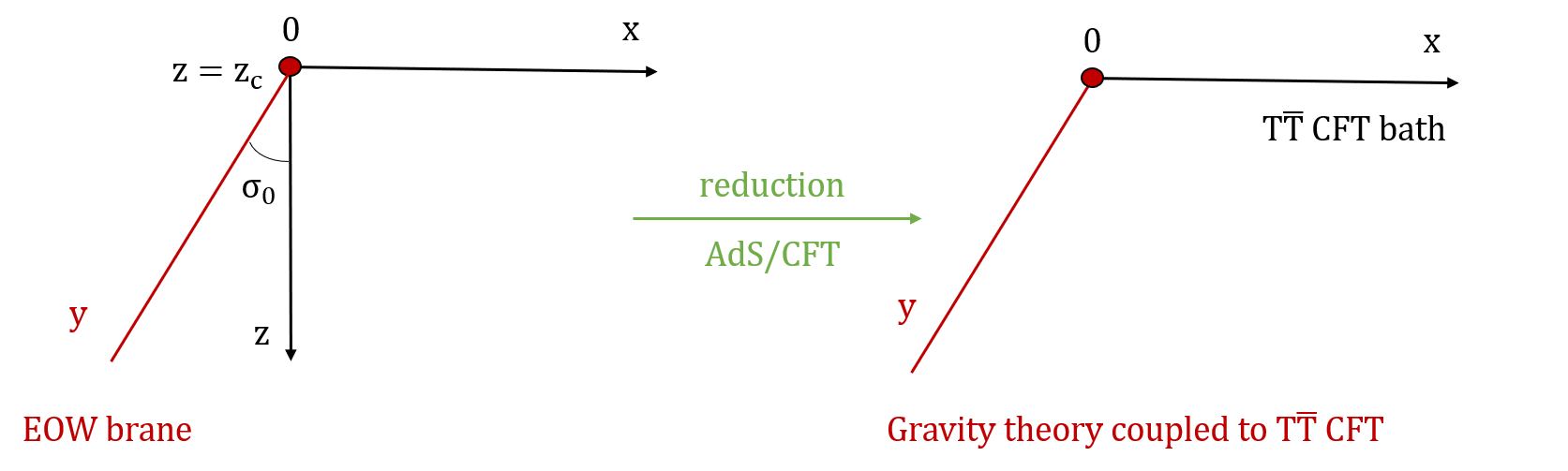}\\
	\caption{$2d$ effective theory obtained by imposing partial reduction in Type I.}\label{red1}
\end{figure}

\begin{figure}
	\centering
	\includegraphics[width=14cm,height=4.5cm]{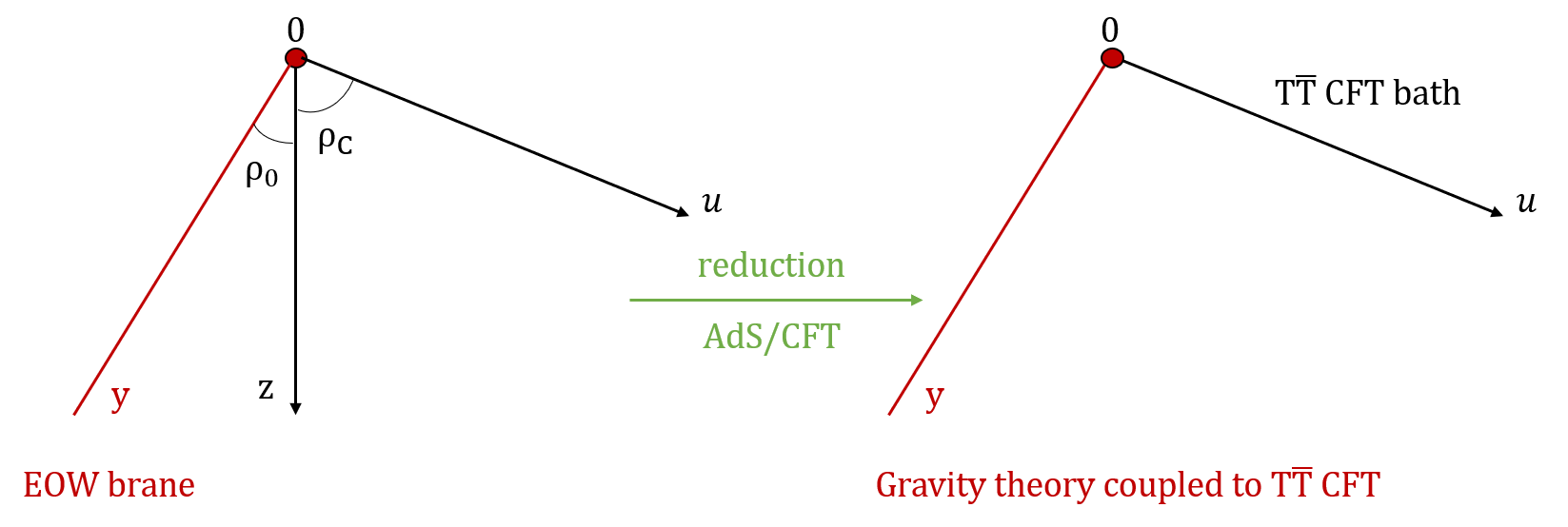}\\
	\caption{$2d$ effective theory obtained by imposing partial reduction in Type II. }\label{red2}
\end{figure}

First let us consider Type I case. The reduction procedure is shown in F.G.\ref{red1}.
Generically, the $2d$ effective Newton constant $G_N^{(2)}$ can be derived from partial reduction by integrating out radial direction for the $3d$ Einstein-Hilbert action. By the formula $\frac{A}{4G_N^{(2)}}$, one can obtain the area term included in the island formula. Equivalently, the $2d$ area term can be considered as (part of) the $3d$ RT surface area. By choosing a point $y=a$ on the EOW brane, the minimal surface will be the RT that ends on the EOW brane $y=a$ and the zero tension brane. We expect that this minimal surface gives the $2d$ area term and it is indeed the case when there is no $T\bar T$ deformation. Thus the area term in the $2d$ effective theory is given by
\begin{equation}
    \begin{split}
        S_{area}(a)=\frac{l}{4G_N} \operatorname{arccosh}\left[\frac{a^2 \tanh ^2\frac{\sigma_0}{l}+a^2+2 a z_c \text{sech} \frac{\sigma_0}{l}+(a\cdot\text{sech}\frac{\sigma_0}{l}+z_c)^2+z_c^2}{2 (a \cdot\text{sech}\frac{\sigma_0}{l}+z_c) \sqrt{a^2+2 a z_c \text{sech}\frac{\sigma_0}{l}+z_c^2}}\right]\ .
    \end{split}
\end{equation}
Let us now consider Type II case. The reduction procedure is shown in F.G.\ref{red2}.
The area term in Type II from partial reduction is simply given by
\begin{equation}
    S_{area}=\frac{\rho_0}{4G_N}\ ,
\end{equation}
which is a constant.

\subsection{Fine-grained entropy}
In this section we compute fine-grained entropy from defect extremal surface and compare it with the result from island formula.
\subsubsection{Defect Extremal Surface}\label{isdes}
Let us first recall the island formula. Consider a system including a gravity theory glued through boundary condition to a non-gravitational quantum field theory (bath). The gravity theory also contains matter sector which is a quantum field theory. The fine-grained entropy for an interval in the bath can be computed by the so called
Island formula \cite{Penington:2019npb,Almheiri:2019psf,Almheiri:2019hni}
\begin{equation}
S=\min _{X}\left\{\operatorname{ext}_{X}\left[\frac{\operatorname{Area}(X)}{4 G_{N}}+S_{\text {semi-cl }}\left(I\cup \text{Rad}\right)\right]\right\}\ ,
\end{equation}
where $X=\partial I$ and $I$  represents the island region. $S_{\text {semi-cl }}\left(I\cup \text{Rad}\right)$ is entanglement entropy in the semi-classical description. By using Island formula, the fine-grained entropy for Hawking radiation can be calculated and it is shown to follow Page curve. Thus Island formula plays important role in resolving black hole information paradox.

In the effort of trying to understand Island formula from holographic point of view, Defect Extremal Surface (DES) is proposed to be the holographic counterpart of Island formula. Let us briefly review Defect Extremal Surface proposal.
By including matter localized on the EOW brane, in \cite{Deng:2020ent}, AdS$_3$/BCFT$_2$ has been improved to contain a defect theory on EOW brane. Due to the quantum entanglement in the matter, one should include the contribution from matter when computing entanglement entropy. Thus the holographic proposal for calculating entanglement entropy (Ryu-Takayanagi prescription) is improved to be the so called Defect Extremal Surface formula
\begin{equation}\label{DESeq}\begin{split}
	S_{\mathrm{DES}}&=\min _{\Gamma}\left\{\operatorname{ext}_{\Gamma,X}\left[\frac{\operatorname{Area}(\Gamma)}{4 G_{N}}+S_{\mathrm{defect}}[D]\right]\right\}\ ,
\end{split}\end{equation}
where $X=\Gamma \cap D$ with $\Gamma$ is the RT surface and $D$ represents the defect brane. $S_{\mathrm{defect}}[D]$ is the entanglement entropy from quantum matter on the defect brane.

To see the relation between Island formula and DES formula, one can calculate fine-grained entropy for the same interval and compare the results. We will do so in the following subsections. We mainly focus on the small $\lambda$ region. In this region one can easily obtain EE in field theory with $T\bar T$ by replacing the UV cutoff by a function of $\lambda$. Let us justify this replacement rule first.

We first consider Type I case. Let us expand the non-perturbative EE (\ref{npe1}) for holographic $T\bar T$ deformed theory in $\lambda\rightarrow 0$ ($z_c\rightarrow 0$) limit\footnote{The perturbative vacuum EE for $T\bar T$ deformed CFT in flat space is obtained in~\cite{Allameh:2021moy} by calculating the replica bulk on-shell action, which is consistent with the RT result.}
\begin{equation}
\begin{split}
       S(A)
       &= \frac{l}{4G_N}\log\frac{2L}{z_c}+
       \frac{l z_c^2}{16G_NL^2}+O(z_c^4)\\
       &= \frac{c}{6}\log(\sqrt{\frac{12}{c \lambda}}2L)+
       \frac{c^2\lambda}{288L^2}+O(\lambda^2)\ .
\end{split}
\end{equation}
It turns out that the leading contribution is nothing but EE in BCFT (with vanishing boundary entropy), $S=\frac{c}{6}\log\frac{2L}{\epsilon}$, with the UV cutoff $\epsilon$ replaced by finite cutoff $\sqrt{\frac{c\lambda}{12}}$. It is believed that in the leading order this replacement gives the correct field theory result, at least for holographic field theory~\cite{Chen:2018eqk}.

For Type II case, the expansion of the non-perturbative EE (\ref{npe2}) in $\lambda\rightarrow 0$ ($\rho_c\rightarrow \infty$) limit is
\begin{equation}
\begin{split}
    S(A)=\frac{\rho_c}{4G_N}=\frac{c}{6}\log(\sqrt{\frac{12}{c\lambda}}2l)-\frac{c^2\lambda}{288l^2}+O(\lambda^2)\ .
\end{split}
\end{equation}
The leading contribution is nothing but EE in AdS$_2$, $S=\frac{c}{6}\log\frac{2l}{\epsilon}$, with the UV cutoff $\epsilon$ replaced by finite cutoff $\sqrt{\frac{c\lambda}{12}}$.
Notice that the finite cutoff is the same in Type I and Type II when expressed in terms of $\lambda$, which is just the cutoff $a(\lambda)=\sqrt{\frac{c\lambda}{12}}$ that we have obtained in section~\ref{II}.
\subsubsection{Type I case}
First let us consider island formula and DES in Type I case.  Here we consider an interval $[0,L]$ in the bath region that contains the boundary. The generalized entropy is given by
\begin{equation}
    \begin{split}
        S_{gen}(a)&=S_{area}(a)+S_{matter}(a,L)\\
        &=\frac{c}{6} \operatorname{arccosh}\left[\frac{a^2 \tanh ^2\frac{\sigma_0}{l}+a^2+2 a z_c \text{sech} \frac{\sigma_0}{l}+(a\cdot\text{sech}\frac{\sigma_0}{l}+z_c)^2+z_c^2}{2 (a \cdot\text{sech}\frac{\sigma_0}{l}+z_c) \sqrt{a^2+2 a z_c \text{sech}\frac{\sigma_0}{l}+z_c^2}}\right]\\
        &+\frac{c}{6}\operatorname{log}\left[\frac{(a+L)^2l}{(z_c+\frac{a}{\cosh{\frac{\sigma_0}{l}}})\epsilon_y z_c}\right]\ ,
    \end{split}
\end{equation}
where $a$ is the boundary position of island and $\epsilon_y=\sqrt{\frac{c\lambda}{12}}$ is the finite cutoff for $T\bar T$ deformed CFT on the brane. According to island formula, the fine-grained entropy is given by minimizing $S_{gen}(a)$. Thus from $\frac{\partial S_{gen}(a)}{\partial a}=0$ we get the extremal point to be
\begin{equation}
    a_{min}=L-2(\cosh{\frac{\sigma_0}{l}}+\sinh{\frac{\sigma_0}{l}})z_c+\mathcal{O}(z_c^2)\ .
\end{equation}
By plugging $a_{min}$ into $S_{gen}(a)$ we get the fine grained entropy for an interval $[0,L]$ in the bath
\begin{equation}
    S_{island}=\frac{c}{6}\frac{\sigma_0}{l}+\frac{c}{6}\operatorname{log}\frac{2L}{z_c}+\frac{c}{6}\operatorname{log}\frac{2l\cosh{\frac{\sigma_0}{l}}}{\epsilon_y}-\frac{c(\cosh{\frac{\sigma_0}{l}}+\sinh{\frac{\sigma_0}{l}})}{6L}z_c+\mathcal{O}(z_c^2)\ .
\end{equation}

Now we compute the fine-grained entropy for the interval $[0,L]$  using Defect Extremal Surface.
The generalized entropy is given by
\begin{equation}
    \begin{split}
        S_{gen}(a)&=S_{brane\ matter}(a)+S_{RT}(a,L)\\
        &=\frac{c}{6}\operatorname{log}\frac{2al}{(z_c+\frac{a}{\cosh{\frac{\sigma_0}{l}}})\epsilon_y}+\frac{c}{6}\operatorname{arccosh}\left[\frac{(L+a\tanh{\frac{\sigma_0}{l}})^2+z_c^2+(z_c+\frac{a}{\cosh{\frac{\sigma_0}{l}}})^2}{2z_c(z_c+\frac{a}{\cosh{\frac{\sigma_0}{l}}})}\right]\ .
    \end{split}
\end{equation}
From $\frac{\partial S_{gen}(a)}{\partial a}=0$ we can determine the extremal point
\begin{equation}
    a_{min}=L-2(\cosh{\frac{\sigma_0}{l}}+\sinh{\frac{\sigma_0}{l}})z_c+\mathcal{O}(z_c^2)\ ,
\end{equation}
which agrees with the extremal point obtained in island case in first order.
Then the fine-grained entropy for an interval $[0,L]$ in the bath from DES is
\begin{equation}
    \begin{split}
        S_{DES}&=S_{gen}(a_{min})\\
        &=\frac{c}{6}\operatorname{log}\left[\cosh{\frac{\sigma_0}{l}}+\sinh{\frac{\sigma_0}{l}}\right]+\frac{c}{6}\operatorname{log}\frac{2L}{z_c}+\frac{c}{6}\operatorname{log}\frac{2l\cosh{\frac{\sigma_0}{l}}}{\epsilon_y}\\
        &-\frac{c(\cosh{\frac{\sigma_0}{l}}+\sinh{\frac{\sigma_0}{l}})}{6L}z_c+\mathcal{O}(z_c^2)\ .
    \end{split}
\end{equation}
This agrees with island result. 
\subsubsection{Type II case}
Now let us compute fine-grained entropy in Type II case. First we use island formula to compute the fine-grained entropy for an interval $[0,L]$. The generalized entropy is given by
\begin{equation}
    \begin{split}
        S_{gen}(a)&=S_{area}(a)+S_{matter}(a,L)\\
        &=\frac{c}{6}\frac{\rho_0}{l}+\frac{c}{6} \operatorname{log}\left[\frac{l(a+L)^2\cosh{\frac{\rho_0}{l}}\cosh{\frac{\rho_c}{l}}}{aL\epsilon_y}\right]\ ,
    \end{split}
\end{equation}
and the extremal point is 
\begin{equation}
    a_{min}=L\ .
\end{equation}
Thus the fine grained entropy obtained from island formula is 
\begin{equation}
\begin{split}
    S_{island}&=S_{gen}(a_{min})\\
        &=\frac{c}{6}\frac{\rho_0}{l}+\frac{c}{6} \operatorname{log}\left[\frac{4l\cosh{\frac{\rho_0}{l}}\cosh{\frac{\rho_c}{l}}}{\epsilon_y}\right]\\
&=\frac{c}{6}\frac{\rho_0}{l}+\frac{c}{6} \operatorname{log}\frac{2l\cosh{\frac{\rho_0}{l}}}{\epsilon_y}+\frac{c}{6}\frac{\rho_c}{l}+\mathcal{O}(e^{-\frac{2\rho_c}{l}})\ .
        \end{split}
\end{equation}
Now we compute the result from DES. The generalized entropy is
\begin{equation}
    \begin{split}
        S_{gen}(a)&=S_{brane\ matter}(a)+S_{RT}(a,L)\\
        &=\frac{c}{6}\operatorname{arccosh}\left[\frac{((L\tanh{\frac{\rho_c}{l}}+a\tanh{\frac{\rho_0}{l}})^2+\frac{L^2}{\cosh^2{\frac{\rho_c}{l}}}+\frac{a^2}{\cosh^2{\frac{\rho_0}{l}}})\cosh{\frac{\rho_0}{l}}\cosh{\frac{\rho_c}{l}}}{2La}\right]\\
        &+\frac{c}{6} \operatorname{log}\frac{2l\cosh{\frac{\rho_0}{l}}}{\epsilon_y}\ ,
    \end{split}
\end{equation}
and the extremal point is determined to be
\begin{equation}
    a_{min}=L\ .
\end{equation}
Thus the fine grained entropy from DES is
\begin{equation}
    \begin{split}
         S_{DES}&=S_{gen}(a_{min})\\
        &=\frac{c}{6}\frac{\rho_0}{l}+\frac{c}{6}\frac{\rho_c}{l}+\frac{c}{6} \operatorname{log}\frac{2l\cosh{\frac{\rho_0}{l}}}{\epsilon_y}\ .
    \end{split}
\end{equation}
One can see that in large $\rho_c$ limit, the two results agree with each other.

\section{Page curve with $T\bar T$ deformed bath}\label{secIV}
In this section, we compute Page curves for evaporating black hole, both Type I and Type II are considered. First we briefly review how to obtain evaporating black hole by doing conformal transformation and analytic continuation. Then we apply the technique to our current model. Finally we find island phase and no-island phase and obtain Page curves. The effects of $T\bar T$ deformation on Page curve are also discussed.
\subsection{Emergence of black hole}
The emergence of black hole from EOW brane can be shown as follows \cite{Chu:2021gdb}. Consider AdS$_3$/BCFT$_2$ in Euclidean space-time
\begin{equation}
    d s^2=l^2 \frac{d \tau^2+d x^2+d z^2}{z^2}\ ,
\end{equation}
we choose the boundary to be $\tau=0$ and BCFT is defined in the region $\tau \geq 0$. The EOW brane in the AdS bulk is located at $\tau=-z\sinh{\frac{\rho_0}{l}}$. Using conformal transformation
\begin{equation}\label{ct}
    \begin{split}
        \begin{aligned}
& \tau=\frac{2l\left(x^{\prime 2}+\tau^{\prime 2}+z^{\prime 2}-l^2\right)}{\left(\tau^{\prime}+l\right)^2+x^{\prime 2}+z^{\prime 2}}\ , \\
& x=\frac{4 x^{\prime}l^2}{\left(\tau^{\prime}+l\right)^2+x^{\prime 2}+z^{\prime 2}}\ , \\
& z=\frac{4 z^{\prime}l^2}{\left(\tau^{\prime}+l\right)^2+x^{\prime 2}+z^{\prime 2}}\ ,
\end{aligned}
    \end{split}
\end{equation}
 the boundary of BCFT is mapped to a circle $x^{\prime 2}+\tau^{\prime 2}=l^2$ and the EOW brane a part of sphere $\left(z^{\prime}+l\sinh{\frac{\rho_0}{l}}\right)^2+x^{\prime 2}+\tau^{\prime 2}=l^2\cosh^2{\frac{\rho_0}{l}} $. There is a black hole on the EOW brane as the horizon can be seen if we do a wick rotation $\tau^{\prime} \rightarrow i t^{\prime}$. With the coordinate transformation
 \begin{equation}
     x^{\prime}=le^{\frac{X}{l}} \cos \phi\ , \quad \tau^{\prime}=le^{\frac{X}{l}} \sin \phi\ ,
 \end{equation}
 the polar coordinate $\phi$ can be identified with the Euclidean time. And the nontrivial time evolution can be seen by wick rotating $\phi$ to physical time $T$
 \begin{equation}
     x^{\prime}=le^{\frac{X}{l}} \cosh \frac{T}{l}, \quad \tau^{\prime}=i le^{\frac{X}{l}} \sinh \frac{T}{l}\ .
 \end{equation}
As before, partial reduction procedure can be used to obtain the $2d$ effective theory, which is an evaporating black hole coupled with CFT bath.
\subsection{Type I case}
First we consider the case with Type I $T\bar T$ deformed CFT.
Notice that in Type I case, the boundary of bath is moved into the bulk and Neumann boundary condition is resolved. Thus the new brane location is $z=\frac{-\tau}{\sinh{\frac{\sigma_0}{l}}}+z_c$. Under conformal transformation (\ref{ct}), the boundary is mapped to a circle $x^{\prime 2}+\tau^{\prime 2}=l^2-{z_c^\prime}^2$ and the EOW brane is still a part of sphere $\left(z^{\prime}+l\sinh{\frac{\sigma_0}{l}}\right)^2+x^{\prime 2}+\tau^{\prime 2}=l^2\cosh^2{\frac{\sigma_0}{l}}+2z^{\prime}_cl\sinh{\frac{\sigma_0}{l}} $, where the relation between $z^{\prime}_c$ and $z_c$ is given by the third equation of (\ref{ct}). Here we consider $z_c^{\prime}$ to be fixed in coordinate $({\tau}^{\prime},x^{\prime},z^{\prime})$.

\begin{figure}
	\centering
	\includegraphics[width=14cm,height=4cm]{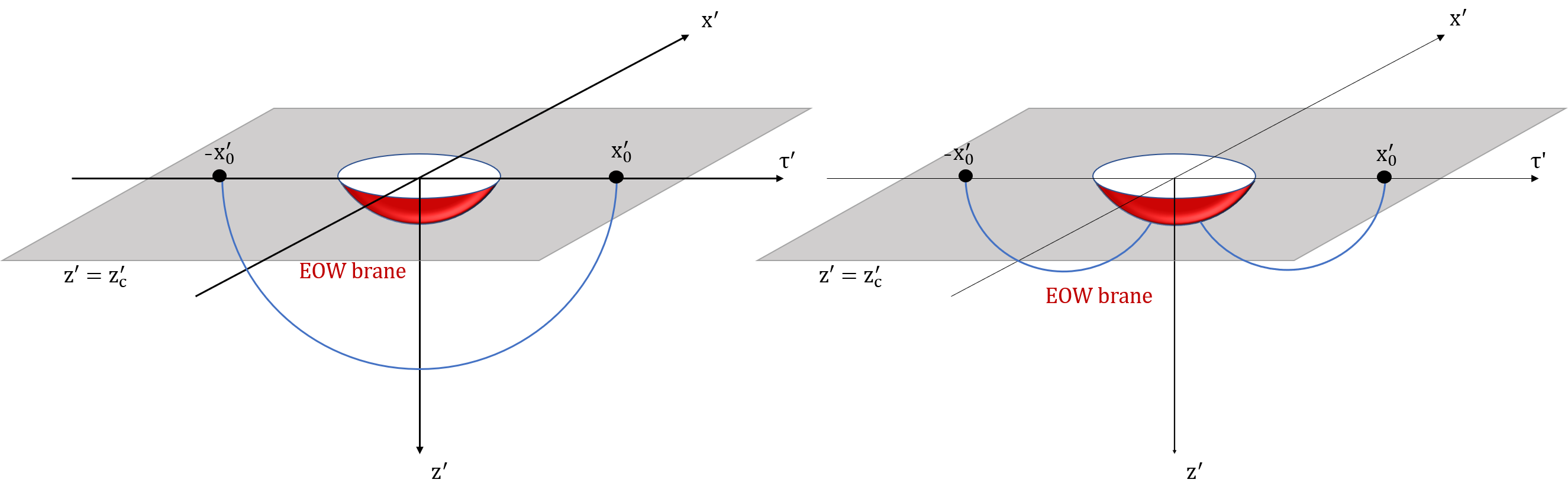}\\
	\caption{Two phases of extremal surfaces in Type I case: connected phase (left) and disconnected (right) phase.}\label{tp1}
\end{figure}

We first use DES to calculate fine-grained entropy for a bath interval $\left[-\infty,-x_0^{\prime}\right] \cup\left[x_0^{\prime}, \infty\right]$ at constant time slice $\tau^{\prime}=\tau_0^{\prime}$. There are two phases of extremal surfaces: connected phase and disconnected phase. Two phases are shown in F.G.\ref{tp1}.

For connected phase, the fine-grained entropy is
\begin{equation}
    S_{DES}=\frac{c}{3} \log \frac{2 x_0^{\prime}}{z^{\prime}_c}+\mathcal{O}({z_c^{\prime}}^2)\ .
\end{equation}

\begin{figure}[h]
  \centering
  \includegraphics[width=9cm,height=5cm]{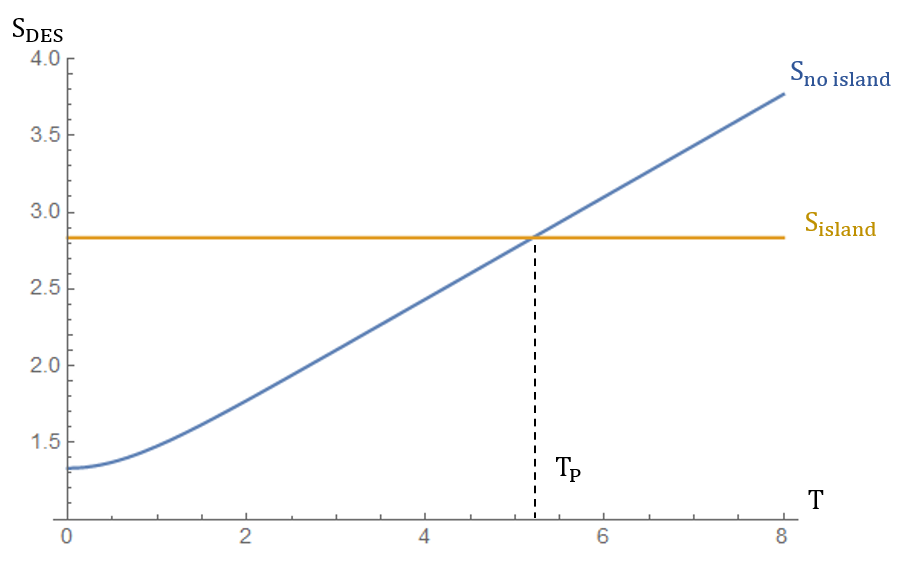}\\
  \caption{Page curves with Type I $T\bar T$ deformation in unit of $c$. We have picked $X_0=1$, $l=1$, $\sigma_0=1$ and $\epsilon_y=z^{\prime}_c=0.1$.}\label{pc1}
\end{figure}

For disconnected phase, we set the intersection points of extremal surfaces and EOW brane to be $\left(\tau_1^{\prime}, \pm x_1^{\prime}, z_1^{\prime}\right)$, or equivalently $\left(-z_1 \sinh{\frac{\sigma_0}{l}}+z_c, \pm x_1, z_1\right)$. Then the generalized entropy is given by
\begin{equation}
    \begin{split}
         S_{gen}&=S_{brane\ matter}+S_{RT}\\
        &=\frac{c}{3}\operatorname{log}\frac{2y_1l}{(z_c+\frac{y_1}{\cosh{\frac{\sigma_0}{l}}})\epsilon_y}\\
        &+\frac{c}{3}\operatorname{arccosh}\left[\frac{(\tau_0+y_1\tanh{\frac{\sigma_0}{l}})^2+(x_0-x_1)^2+z_c^2+(z_c+\frac{y_1}{\cosh{\frac{\sigma_0}{l}}})^2}{2z_c(z_c+\frac{y_1}{\cosh{\frac{\sigma_0}{l}}})}\right]\ ,
    \end{split}
\end{equation}
where $y_1=(z_1-z_c)\cosh{\frac{\sigma_0}{l}}$. By extremizing the generalized entropy, the extremal position is found to be 
\begin{equation}
    \begin{split}
        \left\{\begin{array}{l}
x_1=x_0\\
y_1=\tau_0-2(\cosh{\frac{\sigma_0}{l}}+\sinh{\frac{\sigma_0}{l}})z_c+\mathcal{O}(z_c^2)\ .
\end{array}\right.
    \end{split}
\end{equation}
Thus the fine-grained entropy is given by
\begin{equation}
    \begin{split}
        S_{DES}
        &=\frac{c}{3}\frac{\sigma_0}{l}+\frac{c}{3}\operatorname{log}\frac{2\tau_0}{z_c}+\frac{c}{3}\operatorname{log}\frac{2l\cosh{\sigma_0}}{\epsilon_y}\\
        &-\frac{c(\cosh{\frac{\sigma_0}{l}}+\sinh{\frac{\sigma_0}{l}})}{3\tau_0}z_c+\mathcal{O}(z_c^2)\ .
    \end{split}
\end{equation}
The result can be summarized in $(X,T)$ coordinates (up to $\mathcal{O}({z_c^{\prime}}^2)$)
\begin{equation}
    \begin{split}
        S_{DES}= \begin{cases}\frac{c}{3}\log \frac{2l \cosh \frac{T}{l}}{z^{\prime}_c}+\frac{c}{3}\frac{X_0}{l}\ ,  &T<T_P \\ \frac{c}{3}\log \frac{l^2e^{2 X_0/l}+{z^{\prime}_c}^2-l^2}{z^{\prime}_cl}+\frac{c}{3}\frac{\sigma_0}{l}+\frac{c}{3}\operatorname{log}\frac{2l\cosh{\frac{\sigma_0}{l}}}{\epsilon_y}
        -\frac{2\,l\,c\,e^{\frac{\sigma_0}{l}}z^{\prime}_c}{3(l^2e^{\frac{2X_0}{l}}+{z^{\prime}_c}^2-l^2)}\ , & T>T_P\ .\end{cases}
    \end{split}
\end{equation}
Here $T_P$ is the Page time and is given by
\begin{equation}
    T_P=l\operatorname{arccosh}\left[\frac{(l^2e^{\frac{2X_0}{l}}+{z^{\prime}_c}^2-l^2)\cosh{\frac{\sigma_0}{l}}e^{(\frac{\sigma_0}{l}-\frac{X_0}{l}-\frac{2lz^{\prime}_ce^{\frac{\sigma_0}{l}}}{l^2e^{2X_0/l}+{z^{\prime}_c}^2-l^2})}}{l\epsilon_y}\right]\ .
\end{equation}
It can also be checked straightforwardly that island formula gives the same result, thus the Page curve is obtained in both ways. To see how Type I $T\bar T$ deformation affects black hole evaporation, we differentiate $S_{DES}$ and $T_P$ with respect to ${z_c^{\prime}}$ to observe how the fine-grained entropy of Hawking radiation and the Page time change. We find that\footnote{Notice that here $\epsilon_y=\sqrt{\frac{c\lambda}{12}}=z_c^{\prime}$ which is obtained by plugging into holographic dictionary for $\lambda$ in coordinates with prime.}
\begin{equation}
    \begin{split}
        \frac{\partial S_{DES}}{\partial z_c^{\prime}}= \begin{cases}-\frac{c}{3}\frac{1}{z_c^{\prime}}\ , &T<T_P \\ -\frac{c}{3}\left(\frac{2}{z_c^{\prime}}+\frac{2e^{\frac{\sigma_0}{l}}}{le^{2\frac{X_0}{l}}-l}\right)+\mathcal{O}({z_c^{\prime}})\ , & T>T_P\ .\end{cases}
    \end{split}
\end{equation}
And
\begin{equation}
    \frac{\partial T_P}{\partial z_c^{\prime}}=-\frac{l}{z_c^{\prime}}-\frac{2 e^{\frac{\sigma_0}{l}}}{e^{\frac{2X_0}{l}}-1}+\frac{e^{-2\frac{\sigma_0}{l}}\left(4e^{\frac{2\sigma_0+2X_0}{l}}-4e^{\frac{2\sigma_0}{l}}-e^{\frac{2X_0}{l}}\text{sech}^2{\frac{\sigma_0}{l}}\right)z_c^{\prime}}{2l\left(e^{\frac{2X_0}{l}}-1\right)^2}+\mathcal{O}({z_c^{\prime}})\ .
\end{equation}
Thus the fine-grained entropy will decrease and Page time will be advanced if Type I $T\bar T$  deformation becomes stronger (${z_c^{\prime}}$ becomes bigger). 
The Page curve is plotted in F.G.\ref{pc1}.
\subsection{Type II case}
\begin{figure}
	\centering
	\includegraphics[width=14cm,height=4.5cm]{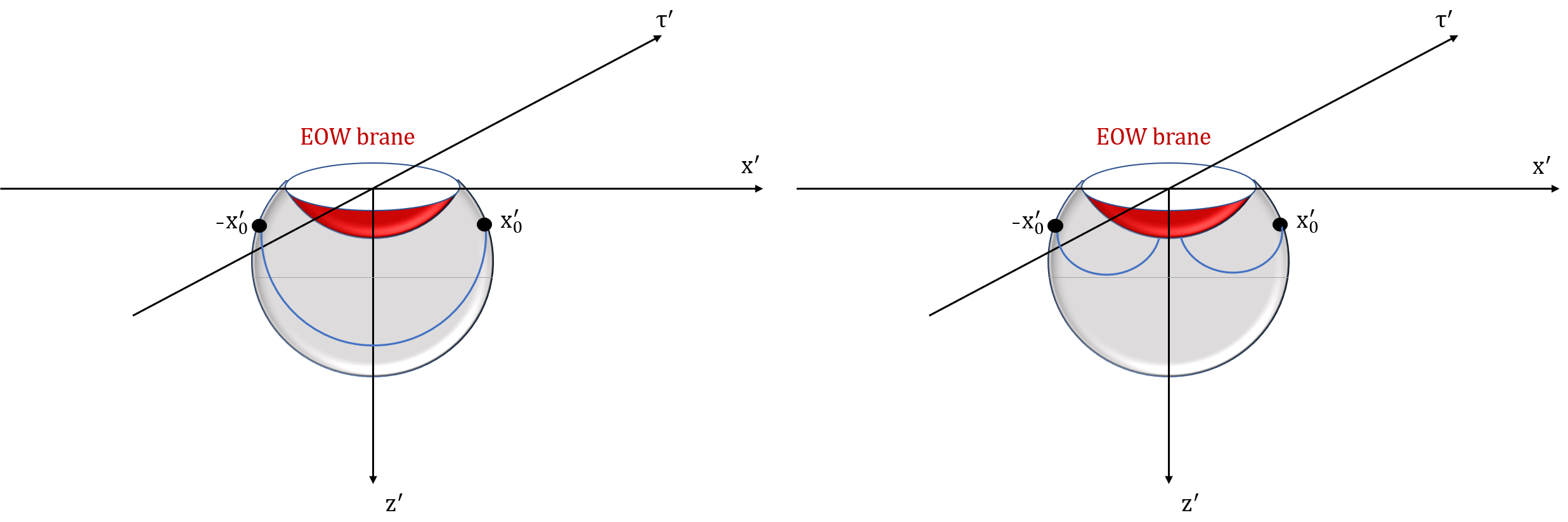}\\
	\caption{Two phases of extremal surfaces in Type II case: connected phase (left) and disconnected (right) phase.}\label{tp2}
\end{figure}

Now we consider Type II $T\bar T$ deformation. The EOW brane stays the same as that without $T\bar T$ deformation whereas the bath transforms to a part of sphere under conformal transformations (\ref{ct}) and is given by $\left(z^{\prime}-l\sinh{\frac{\rho_c}{l}}\right)^2+x^{\prime 2}+\tau^{\prime 2}=l^2\cosh^2{\frac{\rho_c}{l}} $. There are also two phases of extremal surfaces as shown in F.G.\ref{tp2}. The two endpoints of bath interval are $\left(z_0 \sinh{\frac{\rho_c}{l}}, \pm x_0, z_0\right)$.

For connected phase, the fine-grained entropy is
\begin{equation}
    S_{DES}=\frac{c}{6}\operatorname{arccosh}\left[\frac{2{x_0^{\prime}}^2+{z_0^{\prime}}^2}{{z_0^{\prime}}^2}\right]\ .
\end{equation}

For disconnected phase, we set the intersection points of extremal surfaces and EOW brane to be $\left(-z_1 \sinh{\frac{\sigma_0}{l}}, \pm x_1, z_1\right)$, and the generalized entropy is 
\begin{equation}
    \begin{split}
         S_{gen}&=S_{brane\ matter}+S_{RT}\\
        &=\frac{c}{3}\operatorname{arccosh}\left[\frac{((u_0\tanh{\frac{\rho_c}{l}}+y_1\tanh{\frac{\rho_0}{l}})^2+(x_0-x_1)^2+\frac{u_0^2}{\cosh^2{\frac{\rho_c}{l}}}+\frac{y_1^2}{\cosh^2{\frac{\rho_0}{l}}})}{2u_0y_1\operatorname{sech}{\frac{\rho_0}{l}}\operatorname{sech}{\frac{\rho_c}{l}}}\right]\\
        &+\frac{c}{3} \operatorname{log}\frac{2l\cosh{\frac{\rho_0}{l}}}{\epsilon_y}\ ,
    \end{split}
\end{equation}
where $y_1=z_1\cosh{\frac{\rho_0}{l}}$ and $u_0=z_0\cosh{\frac{\rho_c}{l}}$. The extremal condition is 
\begin{equation}
    \begin{split}
        \left\{\begin{array}{l}
x_1=x_0\\
y_1=u_0\ .
\end{array}\right.
    \end{split}
\end{equation}
Thus the fine-grained entropy is 
\begin{equation}
    \begin{split}
         S_{DES}
        =\frac{c}{3}\frac{\rho_0}{l}+\frac{c}{3}\frac{\rho_c}{l}+\frac{c}{3} \operatorname{log}\frac{2l\cosh{\frac{\rho_0}{l}}}{\epsilon_y}\ .
    \end{split}
\end{equation}
The summation of the results is 
\begin{equation}
    \begin{split}
        S_{DES}= \begin{cases}\frac{c}{6}\operatorname{arccosh}\left[\frac{2e^{\frac{2X_0}{l}} \cosh^2 \frac{T}{l}}{(\mu+\sinh{\frac{\rho_c}{l}})^2}+1\right]\ ,  &T<T_P \\ \frac{c}{3}\frac{\rho_0}{l}+\frac{c}{3}\frac{\rho_c}{l}+\frac{c}{3} \operatorname{log}\frac{2l\cosh{\frac{\rho_0}{l}}}{\epsilon_y}\ , & T>T_P\ .\end{cases}
    \end{split}
\end{equation}
And
\begin{equation}
T_P=l\operatorname{arccosh}\left[\frac{e^{\frac{-X_0}{l}}}{\sqrt{2}}\left(\mu+\sinh{\frac{\rho_c}{l}}\right)\left(\cosh\left[\frac{2(\rho_c+\rho_0)}{l}+2\operatorname{log}\frac{2l\cosh{\frac{\rho_0}{l}}}{\epsilon_y}\right]-1\right)^{\frac{1}{2}}\right]\ ,
\end{equation}
where
\begin{equation}
    \mu=\sqrt{\cosh^2{\frac{\rho_c}{l}}-e^{\frac{2X_0}{l}}}\ .
\end{equation}
Next let us see how fine-grained entropy and Page time change under $T\bar T$ deformation. We have\footnote{Here  $\epsilon_y$ equals to the finite cutoff $l/\cosh{\frac{\rho_c}{l}}$. }
\begin{equation}
    \begin{split}
        \frac{\partial S_{DES}}{\partial \rho_c}= \begin{cases}-\frac{ c e^{\frac{X_0}{l}}\cosh \frac{\rho_c}{l}  \cosh \frac{T}{l}}{3 l \mu\left(\mu+\sinh \frac{\rho_c}{l}\right)\left[\frac{ e^{\frac{2 X_0}{l}} \cosh ^2\frac{T}{l}}{\left(\mu+\sinh \frac{\rho_c}{l}\right)^2}+1\right]^{\frac{1}{2}}}<0\ , &T<T_P \\ \frac{c}{3l}(1+\tanh{\frac{\rho_c}{l}})>0\ , & T>T_P\ ,\end{cases}
    \end{split}
\end{equation}
and
\begin{equation}
    \begin{split}
       \frac{\partial}{\partial \rho_c} \frac{\partial S_{DES}}{\partial T}= \begin{cases}-C\left(\sqrt{2} \mu\sinh \frac{\rho_c}{l} +\cosh \frac{2 \rho_c}{l}-e^{\frac{2 X_0}{l}}\right)<0\ , &T<T_P \\ 0\ , & T>T_P\ .\end{cases}
    \end{split}
\end{equation}
where $C$ is a positive number and $\rho_c$ is large.  For the Page time, we have
\begin{equation}
  \frac{\partial T_P}{\partial\rho_c}>0\ .
\end{equation}

\begin{figure}[h]
  \centering
  \includegraphics[width=9cm,height=5cm]{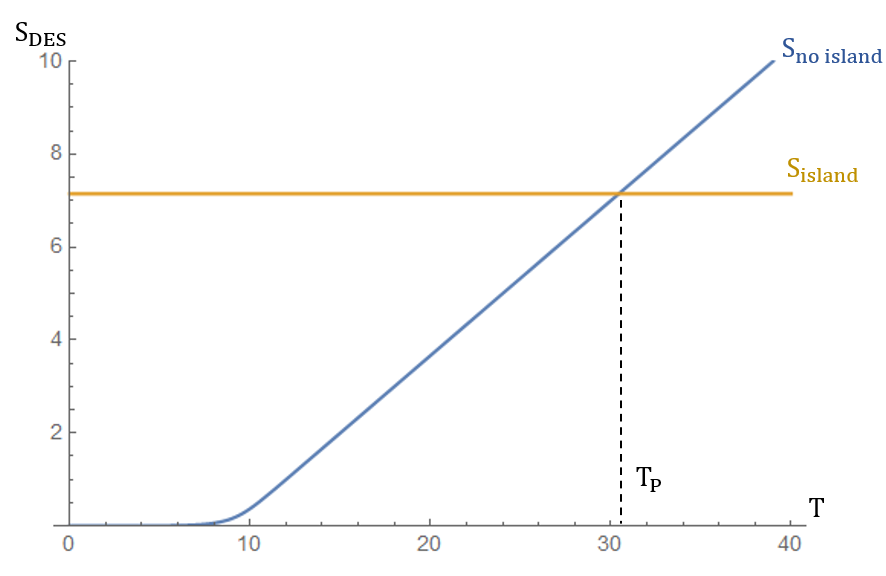}\\
  \caption{Page curve with  Type II $T\bar T$ deformation in unit of $c$. We have picked $X_0=1$, $l=1$, $\rho_0=1$,and $\rho_c=10$.}\label{pc2}
\end{figure}

Thus one can see that as the Type II deformation becomes larger ($\rho_c$ becomes smaller), the black hole will evaporate faster, the fine-grained entropy after Page time will decrease and Page time is advanced. The Page curve is plotted in F.G.\ref{pc2}.
\section{Conclusion and Discussion}\label{secV}
In this paper we study a model of AdS gravity with two different boundaries: one is EOW brane and the other is cutoff boundary. EOW brane in AdS has been extensively studied since the work of Karch and Randall~\cite{Karch:2000ct} and it was holographically interpreted as boundary conformal field theory by Takayanagi~\cite{Takayanagi:2011zk}. In the holographic understanding of recent development about black hole information paradox, in particular the island formula, EOW brane plays important role. A black hole evaporation model can include a brane world gravity plus a quantum field theory bath. This model can be obtained naturally from {\it partial reduction}, which is Karch-Randall reduction for only part of the AdS between finite tension brane and zero tension brane. The reason why partial reduction shows up is that, the bath is the dual of (part of) the bulk and therefore can not interact with the brane world gravity from {\it full reduction} since the latter is the image of itself. A bunch of tests have been done for partial reduction~\cite{Deng:2020ent,Chu:2021gdb,Li:2021dmf,Shao:2022gpg}. Apparently the Neumann boundary condition of EOW brane leads to the reduction to brane world gravity while the Dirichlet boundary condition in the asymptotic boundary requests the AdS/CFT duality. In this paper we develop the idea of partial reduction by considering more general Dirichlet boundaries, namely the finite cutoff surfaces. The field theory interpretation of these boundaries in terms of $T\bar T$ deformation have been recently discussed a lot. We study the model with two different boundaries in great detail and explore many consistent results. In particular we have considered two types of cutoff boundaries, one is half flat space and the other is AdS$_2$. We calculate entanglement entropy in both cases and it agrees with island formula result through partial reduction. We also apply our model to black hole evaporation and find the Page curve with $T\bar T$ deformed bath.

There are a few interesting future questions listed in order: First, it would be interesting to generalize our study to the case where there is a black hole in the bulk. Second, it would be interesting to generalize our study to the case where two boundaries are not connected. Last but not least, it is interesting to apply our model to study holographic cosmology.
\begin{acknowledgments}
We are grateful for the useful discussions with Yunfeng Jiang and Wei Song. This work is supported by NSFC grant 12375063. YZ is also supported by NSFC 12247103 through Peng Huanwu Center for Fundamental Theory. 
\end{acknowledgments}

\bibliographystyle{unsrt}
\bibliography{islandttbar}

\end{document}